\documentclass [11pt]{article}
 \pdfoutput=1
\usepackage{graphicx}
\usepackage{amsmath,amssymb} 
\usepackage{latexsym}
\usepackage{mathrsfs}
\usepackage{bbold}
\usepackage{color}
\usepackage{slashed}
\definecolor{red}{rgb}{1,0,0}
\usepackage{cancel}
\usepackage{comment}

\definecolor{red}{rgb}{1,0,0}

\makeatletter
\def\section{\@startsection {section}{1}{\z@}{-3.5ex plus -1ex minus
 -.2ex}{2.3ex plus .2ex}{\large\bf}}
\def\subsection{\@startsection{subsection}{2}{\z@}{-3.25ex plus -1ex
minus -.2ex}{1.5ex plus .2ex}{\normalsize\bf}}
\makeatother
\makeatletter

\@addtoreset{equation}{section}

\makeatother

\def\bea{\begin{eqnarray}} \def\eea{\end{eqnarray}}
\def\be{\begin{equation}} \def\ee{\end{equation}}

\newcommand{\Lag}{\mathcal{L}}
\newcommand{\Tr}{\text{Tr}}

\newcommand{\SO}{\text{SO}}
\newcommand{\SU}{\text{SU}}
\newcommand{\U}{\text{U}}

\newcommand{\promille}{%
  \relax\ifmmode\promillezeichen
        \else\leavevmode\(\mathsurround=0pt\promillezeichen\)\fi}
\newcommand{\promillezeichen}{%
  \kern-.05em%
  \raise.5ex\hbox{\the\scriptfont0 0}%
  \kern-.15em/\kern-.15em%
  \lower.25ex\hbox{\the\scriptfont0 00}}

\newcommand{\abs}[1]{ \left| #1\right|}

\usepackage[colorlinks,linkcolor=black,citecolor=blue,urlcolor=blue,linktocpage]{hyperref}
\newcommand{\hhref}[2][]{\href{http://arxiv.org/abs/#2#1}{arXiv:#2}}

\begin{document}

\thispagestyle{empty}

\begin{center}

\hfill SISSA 58/2013/FISI \\

\begin{center}

\vspace*{0.5cm}

{\Large\bf  Supersymmetry with a pNGB Higgs and Partial Compositeness}
\end{center}

\vspace{1.4cm}

{\bf David Marzocca$^{a}$, Alberto Parolini$^{a}$ and Marco Serone$^{a,b}$}\\

\vspace{1.2cm}

${}^a\!\!$
{\em SISSA and INFN, Via Bonomea 265, I-34136 Trieste, Italy} 

\vspace{.3cm}

${}^b\!\!$
{\em ICTP, Strada Costiera 11, I-34151 Trieste, Italy}

\end{center}

\vspace{0.8cm}

\centerline{\bf Abstract}
\vspace{2 mm}
\begin{quote}

We study the consequences of combining SUSY with a pseudo Nambu-Goldstone \mbox{boson} Higgs coming from an $\SO(5)/\SO(4)$ coset and ``partial compositeness". In particular, we focus on how electroweak symmetry breaking and the Higgs mass are reproduced in models where the symmetry $\SO(5)$ is linearly realized. The  global symmetry forbids tree-level contributions to the Higgs potential coming from D-terms, differently from what happens in  most of the SUSY little-Higgs constructions. While the stops are generally heavy, light fermion top partners below 1 TeV are predicted. In contrast to what happens in non-SUSY composite Higgs models, they are necessary to reproduce the correct top, rather than Higgs,  mass. En passant, we point out that, independently of SUSY, models where $t_R$ is fully composite and embedded in the ${\bf 5}$ of $\SO(5)$ generally predict a too light Higgs.

\end{quote}

\vfill

\newpage


\section{Introduction}

The discovery of the Higgs boson at the LHC \cite{Aad:2012tfa} and, at the same time, the absence of the discovery of new particles, is becoming a challenge for natural theories, 
aiming to solve the gauge hierarchy problem. This tension applies in particular to supersymmetric (SUSY) models, where the natural scale of new physics beyond the Standard Model (SM) is predicted around the weak scale. 
The ever-increasing bounds on sparticle masses are confining SUSY theories in the per cent or lower region of fine-tuning.

An alternative possible solution to the gauge hierarchy problem is to assume that the Higgs field  is a pseudo Nambu-Goldstone Boson (pNGB) of a spontaneously broken global symmetry.
Since in absence of SUSY scalar masses are unnatural, the obvious framework of this idea is in the context of strongly coupled field theories, where the pNGB Higgs is a bound state
of some more fundamental constituents \cite{Kaplan}, like pions in QCD. In contrast to SUSY, in this scenario new particles can naturally appear at a scale significantly higher than the weak one.
Since the real Higgs is not a NGB, one has to properly add explicit symmetry breaking terms to give it a mass, without reintroducing 
the hierarchy problem.  Moreover, electroweak precision data indicate that the Higgs compositeness scale $f$ has to be somewhat higher than the electroweak scale $v$.

Roughly speaking, the model building with a pNGB Higgs in the last years can be grouped in two different classes.
On one hand, we have little Higgs models \cite{ArkaniHamed:2001nc,ArkaniHamed:2002qx}  where, thanks to an ingenious symmetry breaking mechanism (collective breaking) 
the mild hierarchy between $v$ and $f$ can naturally be realized.
On the other hand, one can give up a dynamical explanation for this splitting and rely on tuning. 
Since the explicit working implementations of the little Higgs idea result in cumbersome models, while  
the tuning to accept in the second case is not very high, we focus in this paper on this second possibility. In the latter models, denoted in what follows as Composite Higgs Models (CHM),
the Higgs potential is typically assumed to be entirely generated at the loop level. 

The recent progress (mostly based on holographic 5D models  \cite{Agashe:2004rs}) revealed that the most successful CHM are those where the SM vectors and fermions are partially composite
due to mass mixing with states of the strongly coupled sector \cite{Kaplan:1991dc,Contino:2006nn}. Due to these mixing, SM vectors and fermions become partially composite;
the lighter the states, the weaker  the mixing. 
Recent studies of CHM with a pNGB Higgs and partial compositeness, where the Higgs potential is calculable, revealed that generically the Higgs is predicted to be heavier than 126 GeV, unless
some fermion resonances are anomalously light \cite{Matsedonskyi:2012ym,Redi:2012ha,Marzocca:2012zn,Pomarol:2012qf}. In particular, parametrizations  of the strongly coupled sector in terms of a single scale ($f$) and a single coupling constant $(g_\rho$)  \cite{Giudice:2007fh} might need to be extended. 

The aim of this paper is to combine SUSY and CHM with partial compositeness. 
Our main motivation is explaining the Higgs mass in a theoretically well-defined and controlled set-up. As well-known, standard minimal SUSY models predict a lighter than 126 GeV Higgs while,
as we have just reminded, CHM tend to give a heavier than 126 GeV Higgs (unless light fermion resonances are assumed). 
It is thus natural to ask what would happen if both scenarios were combined. A more theoretical, but equally important, motivation to pursue this analysis is based on the difficulty to construct a purely non-SUSY
UV completion of CHM, in particular it is challenging to explain the  partial compositeness paradigm.\footnote{See ref.~\cite{Barnard:2013zea} for a recent attempt.}
 On the contrary,  SUSY is of great help in trying to address this question and recently some partial UV completions
of SUSY CHM have been found \cite{Caracciolo:2012je}. 
Since SUSY allows to have technically light scalars, we actually consider models where the Higgs
appears as a pNGB of a spontaneously broken, {\it linearly realized,} $\SO(5)$ global symmetry.
A double protection mechanism is at work to suppress the UV sensitivity of the Higgs mass parameter (SUSY and shift symmetry) \cite{Birkedal:2004xi,Chankowski:2004mq,Berezhiani:2005pb}. In contrast to what happens in more standard models such as the Minimal Supersymmetric Standard Model (MSSM), where the Higgs Vacuum Expectation Value (VEV) is quadratically sensitive to soft mass term parameters, the pNGB Higgs VEV at one-loop level is quadratically sensitive to the wino and bino masses only, and SM superpartners can be decoupled without fine-tuning issues.
We do not specify the whole mechanism of SUSY breaking and parametrize it by soft mass terms in both the elementary and the composite sectors. 

The linear models we consider can be seen as the weakly coupled description of some IR phase of a strongly coupled gauge theory,
where  the Higgs is a composite particle and new vector resonances are expected, or as UV completions on their own,  where the Higgs is elementary up to high scales and no compositeness
occurs. In the former case, we might assume that the theory becomes strongly coupled at relatively low scales, such as $\Lambda=4\pi f$ and determine the low energy values of the non-SM gauge and Yukawa couplings by demanding that 
they all become strong at the scale $\Lambda$.
In the latter case, the absence of new vector resonances  allows to extend the validity of the theory to higher scales, up to around 
$100 f$, above which certain Yukawa couplings reach a Landau pole.\footnote{For simplicity of notation and with some abuse of language,  we denote by  ``elementary" the SM fermions and gauge fields and ``composite" the fields coming from the new exotic sector, where the spontaneous breaking of a symmetry produces the pNGB Higgs, independently of the actual interpretation of the models.
The terms ``elementary sector", ``composite sector" and ``partial compositeness" will also be used.}
We analyze two models, representatives of these two possible interpretations. 

The top quark, key player of ElectroWeak Symmetry Breaking (EWSB), is assumed to be elementary to start with.  
We do not consider here models where the top quark (typically its right-handed component $t_R$) comes directly from the composite sector, since the simplest constructions 
with composite fields in the fundamental representation of $\SO(5)$ predict a too light Higgs.
 As we will briefly show, the reason is independent of SUSY and applies quite generally (in the assumption, of course, of the absence of other large $\SO(5)$ violating parameters not related to the top sector). In contrast to non-SUSY CHM, light fermion top partners are in principle no longer needed to reproduce the correct Higgs mass, because SUSY gives us a new handle.
Two notable mass scales govern the Higgs potential: $f$ and the SUSY breaking soft masses $\widetilde m$.
For  $\widetilde m\gg f$, the models can be seen as a linear completion of the non-SUSY CHM, and the Higgs is expected to be too heavy unless light top partners are present.
On the other hand, for $\widetilde m\ll f$, SUSY is too effective in suppressing the radiatively induced Higgs potential and we get a too light Higgs, independently of the overall mass scale of the exotic particles.
However, light fermion top partners are still predicted. In contrast to what happens in non-SUSY models, they are not linked to the Higgs mass, but rather to the top mass itself\footnote{A similar situation occurs in the holographic CHM of ref.\cite{Pappadopulo:2013vca}.} and to the assumption of perturbativity at least up to the scale $\Lambda \sim 4 \pi f$. In both models, the mass mixing between the top and the fermion resonances is such that the correct top mass is reproduced only if some fermion resonance is around the scale $f$. 
This in particular applies also in the ``strongly coupled" model because of a sort of see-saw mechanism among the fermion resonances that produces a light mass eigenvalue.
Stops can instead be quite heavy, well above the TeV scale. In both models the fine-tuning is around the per cent level or better.
Interestingly enough, the request of having perturbativity up to $\Lambda$, EWSB, correct top and Higgs masses and top partners with electric charge 5/3 above the recent bound found by CMS \cite{Chatrchyan:2013wfa} almost
fix the parameter space of our models.

There are two main differences between our SUSY CHM and the SUSY little Higgs models considered in the past  (see e.g. refs.~\cite{Birkedal:2004xi,Roy:2005hg, Csaki:2005fc,Bellazzini:2008zy, Bellazzini:2009ix} for a partial list of references): i) we accept the mild hierarchy between $v$ and $f$ and\footnote{See ref.~\cite{Craig:2013fga} where a similar approach has recently been advocated in a revival of the SUSY twin Higgs idea \cite{Chacko:2005pe,Falkowski:2006qq,Chang:2006ra}.}  ii) we consider global symmetries associated to orthogonal $(\SO(5))$, rather than unitary, groups.\footnote{See ref.~\cite{Redi:2010yv} for a SUSY model with a pNGB Higgs based on $\SO(5)$.}
Orthogonal groups allow to consider scenarios where there is no D-term tree-level contribution to the whole Higgs potential. This leads to some simplification in the model building.

Linear realizations of CHM with partial compositeness are a useful laboratory where some UV-sensitive observables can be studied in a controlled set-up.
In addition to the prominent example of the Higgs potential itself, one might study for instance the occurrence of the possible large and UV-dependent corrections to the $S$-parameter 
and $Z b\bar b$ coupling, recently pointed out in ref.~\cite{Grojean:2013qca}. As an example of this use, we study how unitarity in $WW$ scattering is recovered in our linear models
and match the results with the more bottom-up approach of ref.~\cite{Contino:2011np}.

The structure of the paper is as follows. In section~\ref{Sec:GeneralSetUp} we discuss the general set-up underlying our models and describe the features of the Higgs potential. 
In sections \ref{sec:tRE} and \ref{sec:tREVect} two concrete models, without and with vector resonances, respectively, are introduced.
We conclude in section \ref{conclusions}. In appendix \ref{app:delta} we give some technical details on the parametrization of the Higgs potential, while in appendix \ref{WWscatte} we report our results for the unitarization of $WW$ scattering.

\section{General Set-up}
\label{Sec:GeneralSetUp}

\begin{figure}[!t]
\begin{center}
		\includegraphics[width=110mm]{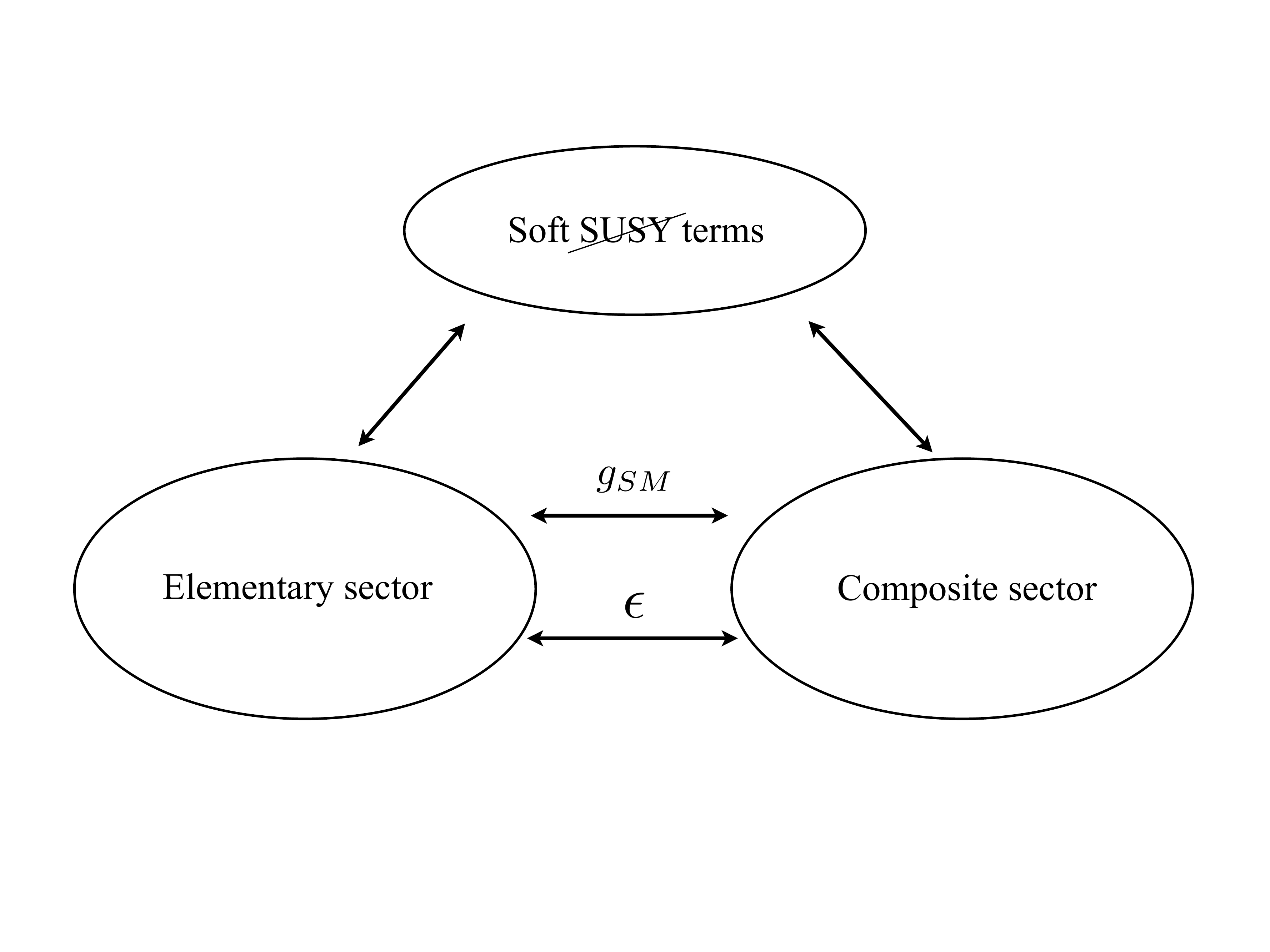}
\end{center}
\vspace*{-2cm}
\caption{\label{fig:STNr2Na1Ns1Nq1_HH} 
\small
Schematic representation of the structure of our models.
}
\label{fig1}
\end{figure}
Our models consist of an elementary sector, containing SM fermions, gauge bosons and their supersymmetric partners, coupled to a composite sector where both the global symmetry $G$ and SUSY are spontaneously broken.  On top of this structure, in order to have sizable SM soft mass terms,  we need to assume the existence of  a further sector which is responsible for an additional source of  SUSY breaking and its mediation to the other two sectors. We do not specify it and we parametrize its effects by adding soft terms in both the elementary and the composite sectors. 
Our key assumption is that the soft masses in the composite sector are $G$ invariant. See fig.\ref{fig:STNr2Na1Ns1Nq1_HH} for a schematic representation. 
The main sources of explicit breaking of $G$ are the couplings between the elementary and the composite sectors, namely the SM gauge couplings and the top mass mixing terms. We assume that partial compositeness in the matter sector is realized through a superpotential portal of the form
\be
W\supset \epsilon \, \xi_{SM} N_{comp}\,.
\label{mH6}
\ee
In eq.(\ref{mH6})  $N_{comp}$ are chiral fields in the composite sector and $\xi_{SM}$ denote the SM matter chiral fields. No Higgs chiral fields  are present
in the elementary sector,  since the Higgs arises from the composite sector. The term (\ref{mH6}) is the only superpotential term involving SM matter fields. 
For concreteness, we consider in this paper only the minimal custodially invariant $\SO(5)\rightarrow \SO(4)$ symmetry breaking pattern 
with $N_{comp}$  in the fundamental representation of $\SO(5)$. 
Like in non-SUSY CHM, the SM Yukawa couplings arise from the more fundamental proto-Yukawa couplings of the form (\ref{mH6}).\footnote{In the field basis where we remove non-derivative interactions of the pNGB Higgs from the composite sector, the Higgs appears in eq.(\ref{mH6}).} 
We do not consider SM fermions but the top in this paper,  since they are not expected to play an important role in the EWSB mechanism.
They can get a mass via partial compositeness through the portal (\ref{mH6}), like the top quark,  or by irrelevant deformations, for instance by adding quartic superpotential terms. 

As mentioned in the introduction, the SUSY models we consider can be seen as the weakly coupled description of some IR phase of a strongly coupled theory, in which case the Higgs is really composite, or alternatively one can take them as linear UV completions, in which case no compositeness occurs.
Depending on the different point of view, general considerations can be made. If we want to take our models as UV completions on their own, we might want to extend the range of validity of the theory up to high scales, ideally up to the GUT or Planck scale. In this setting, introducing gauge fields in addition to the SM gauge fields is disfavoured, because the multiplicity of the involved fields would typically imply that the associated gauge couplings are not UV free and develop a Landau pole at relatively low energies. Avoiding analogous Landau poles for certain Yukawa couplings in the superpotential implies that the ``composite sector" should be as weakly coupled as possible. However, reproducing the correct top mass forces some coupling to be sizable; in our explicit example a Landau pole is reached at a scale around $10^2f$.
Viceversa, additional gauge fields are generally  required if we assume that the linear models considered are an effective IR description of a more fundamental strongly coupled theory, like in ref.~\cite{Caracciolo:2012je}. 
We might now assume that the theory becomes strongly coupled at relatively low scales, such as $\Lambda=4\pi f$. We can actually determine the low energy non-SM Yukawa and gauge couplings by demanding that  they all become strong around the same scale $\Lambda$. As we will see, light fermion top partners still appear in both cases.

In light of these two different perspectives, we will consider in the next sections two benchmark models, with and without vector resonances.


\subsection{Features of the Higgs Potential}

\label{sec:HiggsMass}

When the Higgs is a pNGB associated to an approximate spontaneous symmetry breaking, its VEV is effectively an angle. 
For this reason it is  often convenient to describe its potential not in terms of the Higgs field $h$ itself,\footnote{In order to simplify our notation considerably, we work throughout the paper in the unitary gauge and denote by $h$ the Higgs field in this gauge.}  but of its sine:
\be
s_h \equiv \sin \frac{h}{f}\,,
\label{shDef}
\ee
where $f$ is the Higgs decay constant.
Following a standard notation we also define 
\be
\xi \equiv \langle s_h^2 \rangle \,.  
\ee
The electroweak scale is fixed to be $v^2=f^2 \xi\simeq (246\,{\rm GeV})^2$.
 We focus on small values of $\xi$ and in explicit results we set it to the benchmark value $\xi=0.1$.
Due to the contribution of particles whose masses vanish for $s_h=0$ (such as the top, $W$ and $Z$), the one-loop Higgs potential contains non-analytic terms of the form $s_h^4 \log s_h$ that  do
not admit a Taylor expansion around $s_h=0$. In the phenomenological regions of interest, these terms do not lead to new features and are qualitatively but {\it not} quantitatively negligible.
However, they make an analytic study of the potential slightly more difficult. For this reason, we neglect them altogether in what follows
and refer to the appendix \ref{app:delta} for a more refined analysis of the Higgs potential where they are included.
For $s_h\ll 1$, the tree-level + one-loop potential  $V=V^{(0)}+V^{(1)}$ admits an expansion of the form
\be
V = - \gamma s_h^2 +\beta s_h^4 +{\cal O}(s_h^6)\,.
\label{PotExpdelta}
\ee
The non trivial minimum of the potential is found at
\be
\xi = \frac{\gamma}{2\beta}\,,
\label{xi0}
\ee
and the Higgs mass square is given by 
\be
M_H^2 =  \frac{8 \beta}{f^2} \xi (1-\xi)\,. 
\label{mH0}
\ee
In all the models we will consider, there are two distinct sectors that do not couple at tree-level at quadratic order: a ``matter" sector, including the fields that mix with the top quark and
a ``gauge" sector, including the SM gauge fields and other fields, neutral under color.
The matter and gauge sectors contribute separately to the one-loop Higgs potential (\ref{PotExpdelta}):
\be\begin{split}
\gamma & = \gamma_{gauge} + \gamma_{matter}\,, \\
\beta & = \beta_{gauge} + \beta_{matter}\,.
\end{split} \label{bgdGen}
\ee
The explicit $\SO(5)$ symmetry breaking parameters are the SM gauge couplings $g$ and $g^\prime$ in the gauge sector and
the mixing parameter $\epsilon$ given by eq.(\ref{mH6}) in the matter sector. Since the latter is sizably larger than the former, for sensible values of the parameters 
$\beta_{matter}\gg \beta_{gauge}$.\footnote{A numerical analysis confirms this  result  and shows that typically $\beta_{gauge}$ is at least one order of magnitude smaller than $\beta_{matter}$.}
At fixed $\xi$, then, the Higgs mass is essentially determined by the matter contribution 
(in the numerical study, however, we keep all the contributions to the one-loop potential).
The gauge contribution should instead be retained in $\gamma$, because the fine-tuning cancellations needed to get $\xi \ll 1$, might involve $\gamma_{gauge}$.

\subsection{Non-SUSY Higgs Mass Estimates}
 
Before analyzing the Higgs potential in SUSY CHM, it might be useful to quickly review the 
situation in the purely non-SUSY bottom-up constructions. We focus  in what follows on models where the
composite fields are in the fundamental representation of  $\SO(5)$. Higher representations lead to a multitude of other fields, they are more complicated to embed in a UV model and worsen the problem of 
Landau poles. Moreover they might lead to dangerous tree-level Higgs mediated flavor changing neutral currents \cite{Agashe:2009di}. It should however be emphasized that they can be useful and can result in qualitatively different results, see e.g.  ref.~\cite{Panico:2012uw} for a recent discussion of the Higgs mass estimate for CHM with composite fermions in the ${\bf 14}$ of $\SO(5)$.
Generically, the Higgs mass is not calculable in CHM,
since both $\gamma$ and $\beta$ defined in eq.(\ref{PotExpdelta}) are divergent and require a counterterm.
The situation improves if a symmetry, such as collective breaking \cite{Panico:2011pw,DeCurtis:2011yx}, is advocated to protect these quantities, at least at one-loop level,
or if one assumes that $\gamma$ and $\beta$ are dominated by the lightest set of resonances in the composite sector, saturating generalized Weinberg sum rules  \cite{Marzocca:2012zn,Pomarol:2012qf}, 
in close analogy to what happens in QCD. As far as the Higgs mass is concerned, we see from eq.(\ref{mH0}) that, at {\it fixed} $\xi$, it is enough to make
$\beta$ finite to be able to predict the Higgs mass.

In CHM with partial compositeness, the largest source of explicit breaking of the global symmetry comes from the mass term mixing the top with the composite sector.
In first approximation, we can switch off all other sources of breaking, including the electroweak SM couplings $g$ and $g^\prime$. 
The estimate of the Higgs mass is then necessarily linked to the mechanism generating a mass for the top. 
Let us first consider the case in which both $t_L$ and $t_R$ are elementary.
In this case two mass mixing terms $\epsilon_L$ and $\epsilon_R$ are required to mix them with fermion states of the composite sector.
The top mass goes like
\be
M_{top} \sim \frac{\epsilon_L \epsilon_R}{M_f} s_h\,,
\label{mH4}
\ee
where by $M_f$ we denote the mass (taken equal for simplicity) of the lightest fermion resonances in the composite sector that couple to $t_R$ and $t_L$.
In the limit in which $\epsilon$ is the only source of explicit symmetry breaking, a simple NDA estimate gives the form of the factors $\gamma$, $\beta$ entering in the Higgs potential (\ref{PotExpdelta}):\footnote{The estimate 
(\ref{mH2}) changes when fields in higher representations are considered. For instance, $\beta \sim \frac{N_c M_f^2 \epsilon^2}{16\pi^2}$ when fields in the ${\bf 14}$ are considered \cite{Panico:2012uw}.}
\be
\gamma \sim \frac{N_c}{16\pi^2} \epsilon^2 M_f^2 \,, \quad
\beta \sim \frac{N_c }{16\pi^2} \epsilon^4 \,,
\label{mH2}
\ee
where $N_c=3$ is the QCD color factor. Plugging eqs.(\ref{mH2}) and (\ref{mH4}) in eq.(\ref{mH0}) gives
\be
M_H^2 \simeq \frac{N_c \epsilon^4}{2\pi^2 f^2} \xi \sim  \frac{N_c}{2\pi^2} M_{top}^2 \frac{M_f^2}{f^2} \ \ \ \ \ \ (t_R \;\; {\rm elementary})\,.
\label{mH5}
\ee
This estimate reveals a growth of the Higgs mass with the top partners mass scale. If one assumes that the composite sector is characterized by the single coupling constant $g_\rho$ \cite{Giudice:2007fh}, we expect that $M_f \simeq g_\rho f$. Indirect bounds on the S parameter require $g_\rho f\gtrsim$ 2 TeV. 
For values of $f\lesssim 1$ TeV this implies $g_\rho \gtrsim 2$. In many explicit models \cite{Matsedonskyi:2012ym,Redi:2012ha,Marzocca:2012zn,Pomarol:2012qf} it has been shown that such a choice results in a too heavy Higgs. Indeed, a 126 GeV Higgs is attained only if one assumes that another mass scale characterizes the composite sector and one has relatively light fermion resonances in the composite sector, with $M_f < g_\rho f$. Although the splitting required between $M_f$ and $g_\rho f$ is modest, it is not easy to argue how it might appear in genuinely strongly coupled non-SUSY theories.

Another possibility is having $t_L$ elementary and $t_R$ fully composite. We can now have a direct mixing between $t_L$ and $t_R$, in principle with no need of  composite resonances, that can all be taken heavy.
Denoting by $\epsilon$ this mass mixing term, we get 
\be
M_{top}\simeq \epsilon s_h  \,.
\label{mH1}
\ee
Proceeding as before, we get
\be
M_H^2 \simeq \frac{N_c \epsilon^4}{2\pi^2 f^2} \xi =  \frac{N_c}{2\pi^2} M_{top}^2 \frac{M_{top}^2}{v^2} \ \ \ \ \ \ (t_R \;\; {\rm composite})\,.
\label{mH3}
\ee
We see that the Higgs mass is at leading order independent of the details of the composite sector and tends to be too light.\footnote{
The problem of  a too light Higgs when $t_R$ is fully composite (when embedded in a ${\bf 5}$ of $\SO(5)$) was already pointed out in \cite{Marzocca:2012zn}, where a formula like eq.(\ref{mH3}) (see eq.(5.14)) was derived for a particular model.}
Of course,  this is the case in the assumption that the top mixing term is the dominant source of explicit $\SO(5)$ breaking.
One can always add extra breaking terms to raise the Higgs mass.
Clearly, this is quite ad hoc, unless these terms are already present for other reasons.
This happens in the concrete model with composite $t_R$ introduced in \cite{Caracciolo:2012je}, where anomaly cancellation and absence of massless non-SM states require adding 
exotic elementary states that necessarily introduce an extra source of explicit $\SO(5)$ breaking.
We have explicitly verified in the model  of \cite{Caracciolo:2012je}  that the estimate (\ref{mH3}) captures to a good accuracy the top contribution to the Higgs mass.  This is still too light, despite the presence of additional sources of $\SO(5)$ breaking,
that cannot be taken too large for consistency. 
We conclude that models with a composite $t_R$, at least those where the top sector plays a key role in the EWSB pattern,  lead to a too small Higgs mass.
For this reason, we will not consider in this paper models where $t_R$ is fully composite and only
focus on the case where both $t_L$ and $t_R$ are elementary.

Let us now briefly mention on how $\xi$ can be tuned to the desired value. There are essentially two ways to do that in a calculable manner: either $|\gamma_{matter}| \gg |\gamma_{gauge}|$, in which case the cancellation takes place mostly inside the matter sector, or $|\gamma_{matter}| \sim |\gamma_{gauge}|$, so that the gauge and matter contributions can be tuned against each other  (see, for example, the discussion in Section 4 of \cite{Marzocca:2012zn}). Both options are generally possible, with the exception of minimal (i.e. where $\epsilon$ is the only source of SO(5) violation in the matter sector) models with a composite $t_R$ embedded in a fundamental of $\SO(5)$, where one can rely only on the second option.

\subsection{Higgs Potential in SUSY Models}

Let us now consider more specifically the Higgs potential in SUSY models.  As we mentioned in the introduction, no tree-level D-term contribution to the potential is present
in our models, in contrast to many SUSY little Higgs constructions. 
The latter are based on global unitary symmetries, where one typically embeds the two MSSM Higgs doublets in two distinct multiplets of the underlying global symmetry group. Because of that,  
one generally ends up in having too big D-term contributions to the Higgs mass, whose cancellation usually requires some more model building effort.
In our case, instead, the two Higgs doublets are embedded in a single chiral field $q_{\bf 4}$ that is in the ${\bf 4}$ of the unbroken $\SO(4)$ group.
More precisely, the two Higgs doublets $H_{u,d}$  are embedded in $q_{\bf 4}$ as follows:
\be
q_{\bf 4} =\frac{1}{\sqrt{2}} \bigg(-i({H_u^{(u)}+H_d^{(d)}),H_u^{(u)}-H_d^{(d)},i(H_u^{(d)}-H_d^{(u)}),H_d^{(u)}+H_u^{(d)}}\bigg)\,,
\ee
where the superscript denotes the up or down component of the doublet.
Thanks to the underlying global symmetry, the $H_u$ and $H_d$ soft mass terms are equal, thus $v_u=v_d$ and $\tan \beta = 1$. 
The mass eigenstates are simply the real and imaginary components of $q_{\bf 4}$. Im$\,q_{\bf 4}$ is identified with the heavy Higgs doublet, while Re$\,q_{\bf 4}$ 
is the light (SM) Higgs doublet.
No D-term contribution affects the Higgs mass. In fact, at tree-level the SM Higgs is massless  and its VEV is undetermined. 
Of course, the situation changes at one-loop level, because of the various  sources of violation of the $\SO(5)$ global symmetry. The SM Higgs will still sit along the flat direction (i.e. $\tan \beta$ remains one at the quantum level), but
quantum corrections will lift the flat direction, fix its VEV and give it a mass. As explained at the beginning of subsection \ref{sec:HiggsMass}, being the light Higgs doublet a pNGB, it is convenient to parametrize its potential in terms
of the sine of the field, as in eq.(\ref{shDef}). From now on, for simplicity,  we denote the SM light Higgs doublet as the Higgs and denote by $h$ the Higgs field in the unitary gauge, matching the notation
with that introduced at the beginning of the subsection \ref{sec:HiggsMass}.

In the Dimensional Reduction (DRED) scheme the one-loop Higgs potential $V^{(1)}$ is given by
\be\begin{split}
	V^{(1)}(s_h) & = \frac{1}{16\pi^2}\sum_n \frac{(-1)^{2 s_n} }{4} (2 s_n + 1) m_n(s_h)^4 \left( \log \frac{m_n^2(s_h)}{Q^2} - \frac{3}{2} \right) \\
	& = \frac{1}{64\pi^2} \text{STr}\left[ M^4(s_h) \left( \log\frac{M^2(s_h)}{Q^2} - \frac{3}{2} \right) \right],
	\label{eq:1loop_pot}\end{split}
\ee
where $m^2_n(s_h)$ are the Higgs-dependent mass squared eigenvalues for the scalars, fermions and gauge fields in the theory and we have denoted the sliding scale by $Q$.
When the mass eigenvalues are not analytically available, we compute the $\log M^2$ term by using the following identity, valid for an arbitrary semi-positive definite matrix $M$, see e.g. ref.~\cite{Shih:2007av}:
\be
M^4 \log M^2 =\lim_{\Lambda\rightarrow \infty} \Big(\frac 12 \Lambda^4 - \Lambda^2 M^2 + M^4 \log \Lambda^2 - 2 \int_0^\Lambda \frac{x^5dx}{x^2+M^2}\Big)\,.
\label{M4logM2}
\ee
The RG-invariance of the scalar potential at one-loop level reads
\be
	\frac{\partial}{\partial \log Q}  V^{(1)} + \beta_{\lambda_I} \frac{\partial}{\partial\lambda_I} V^{(0)} -\gamma_n \Phi_n  \frac{\partial}{\partial \Phi_n} V^{(0)} = 0,
	\label{eq:scale_dependence}
\ee
where the indices $I$ and $n$ run over all the masses and couplings (including soft terms) and all the scalar fields in the theory, respectively, and $V^{(0)}$ denotes
the tree-level scalar potential with the addition of soft terms.
By expanding eq.~(\ref{eq:1loop_pot}) for $s_h\ll 1$, we get the explicit form for $\gamma$ and $\beta$ defined in eq.~(\ref{PotExpdelta}).
As we already pointed out in section \ref{sec:HiggsMass}, in first approximation
we can switch off all SM gauge interactions and keep only the top mixing masses $\epsilon$ as explicit source of symmetry breaking.
In this limit, only colored fermion and scalar fields contribute to the Higgs potential (\ref{eq:1loop_pot}).

When all sources of SUSY breaking, denoted collectively by  $\widetilde m$, are switched off, SUSY requires
\be
\lim_{\widetilde m\rightarrow 0} V(h) = 0\,.
\label{Vsusy0}
\ee
However, one has to be careful in properly taking the two limits $\widetilde m\rightarrow 0$, and $s_h\rightarrow 0$, since in general they do not commute.
The cancellation (\ref{Vsusy0}) is only manifest when we first take the $\widetilde m\rightarrow 0$ limit. In practice, however, we only expand in $s_h$ since the sources of SUSY breaking cannot be taken too small.
 
When the soft terms in the composite sector are $\SO(5)$ invariant and the SM gauge interactions are switched off, so that the only $\SO(5)$ violating term is the superpotential term (\ref{mH6}), the $\beta$-functions $\beta_{\lambda_I}$ and the anomalous dimensions $\gamma_n$
appearing in eq.(\ref{eq:scale_dependence}) are necessarily $\SO(5)$ invariant at one-loop level. As a consequence, neither the second nor the third term in eq.~(\ref{eq:scale_dependence}) can
depend on $s_h$ and hence the $s_h$-dependent one-loop potential $V^{(1)}$ is RG invariant and finite. 
In this case, in contrast to the MSSM, the electroweak scale $\xi$ defined in eq.(\ref{xi0}) is only logarithmically sensitive to the soft masses when  these are taken parametrically large. 
The global symmetry breaking scale $f$ is quadratically sensitive to the soft mass terms associated to the fields responsible for this breaking when these are taken parametrically large.
In our models such fields are always in the gauge sector, where we provide a dynamical mechanism of SUSY breaking. For this reason and for simplicity, we do not introduce composite soft mass terms in the gauge sector.

When the SM gauge interactions are switched on,
$\beta_{\lambda_I}$ and $\gamma_n$ are no longer $\SO(5)$ invariant and can depend on $s_h$.
Although holomorphy protects the superpotential from quantum corrections, the K\"ahler potential is renormalized and the gauging
of $\SU(2)_L\times \U(1)_Y$ explicitly breaks the $\SO(5)$ global symmetry. This implies that  the physical, rather than holomorphic, couplings of the composite sector entering in the superpotential split into several components
with different RG evolutions, depending on the $\SU(2)_L\times \U(1)_Y$ quantum numbers of the involved fields.  In what follows, we take the physical couplings to be all equal at the scale $f$.
Similarly, the RG flow induced by the SM gauge couplings gives rise
to $\SO(5)$ violating contributions to the soft mass terms. In the models we will consider this dependence appears only at order $s_h^2$. 
It implies that the RG flow of the tree-level soft terms contribute to $\gamma$
and induce a quadratic sensitivity to the wino and bino soft terms suppressed by a one-loop factor $\sim g^2/(16\pi^2)$.
A ``Higgs soft mass term'' of the form $\frac{1}{2} \widetilde m_H^2 f^2 s_h^2$,  even if absent at tree-level, is radiatively generated by the bino and wino masses $\widetilde m_g$.
A radiatively stable assumption about the Higgs soft term $\widetilde m_H^2$ is to take it at the scale $f$ of order 
\be
	| \widetilde{m}_H^2 | \sim \frac{g^2}{16 \pi^2} \widetilde{m}_g^2 \ . 
	\label{eq:boundmHSQ}
\ee
In this way, we can neglect its effect on the one-loop potential.


\section{Minimal $\SO(5) \rightarrow \SO(4)$ Model}

\label{sec:tRE}

A simple supersymmetric pNGB Higgs Model  with elementary $t_L$ and $t_R$ can be constructed using two colored chiral multiplets $N_{L,R}$ in the ${\bf 5}$ of $\SO(5)$,
two colored $\SO(5)$ singlet fields $S_{L,R}$, two color-neutral multiplets in the ${\bf 5}$, $q$ and $\psi$, and a complete singlet $Z$. All these multiplets are necessary to have a linear realization of the global symmetry breaking
$\SO(5)\rightarrow \SO(4)$ without unwanted massless charged states. The superpotential reads
\be
W = \sum_{i=L,R} (\epsilon_i \xi^a_i N_i^a + \lambda_i S_i q^a N_i^a )+ m_N N_L^a N_R^a + m_S  S_L S_R+ W_0(Z,q,\psi)\,,
\label{WtRC}
\ee
where 
\be
W_0(Z,q,\psi) = h Z (q_a q_a - \mu^2) + m_\psi q_a \psi_a\,.
\label{W0tRel}
\ee
We embed the elementary $q_L$ and $t_R$ into spurions $\xi_L$ and $\xi_R$ in the ${\bf 5}$ of $\SO(5)$ for minimality:\footnote{In order to keep the notation light, 
we omit in what follows the color properties of the fields, that should be clear from the context.}
\begin{equation}
\xi_L=\frac{1}{\sqrt{2}}\left(\begin{array}{c}
b_{L}\\
-ib_{L}\\
t_{L}\\
it_{L}\\
0\end{array}\right)\,, \qquad
\xi_R=\left(\begin{array}{c}
0\\
0\\
0\\
0\\
t_R\end{array}\right)\,.
\label{EletR2}
\end{equation}
The superpotential (\ref{W0tRel}) corresponds to an O'Raifeartaigh model of SUSY breaking. For $\mu^2> m_\psi^2/(2h^2)$, this model has a SUSY breaking minimum with 
\be
\langle q_a \rangle = \frac{f}{\sqrt{2}} \delta_a^5\,,
\label{qVEVelenovect}
\ee
where
\be
f = \sqrt{2\mu^2-\frac{m_\psi^2}{h^2}}\,.
\label{fDEF}
\ee
The scalar VEV's of $Z$ and $\psi_a$, undetermined at the tree-level, are stabilized at the origin by a one-loop potential.
The symmetry breaking pattern is the minimal
\be
\SO(5)\times \U(1)_X \rightarrow \SO(4)\times \U(1)_X \,,
\ee
where $\SU(2)_L\times \U(1)_Y$ is embedded in $\SO(4)\times \U(1)_X$ in the standard fashion.
The four NGB's $h^{\hat{a}}$ can be described by means of the $\sigma$-model matrix as 
\be
	q_a = U_{ab} \tilde q_b  = \exp\left( \frac{i \sqrt{2}}{f} h^{\hat{a}} T^{\hat{a}}  \right)_{ab} \tilde q_b,
	\label{mH10}
\ee
where $T^{\hat a}$ are the $\SO(5)/\SO(4)$ broken generators defined as in the Appendix A of ref.~\cite{Caracciolo:2012je} and $\tilde q$ encodes the non-NGB degrees of freedom of $q$.
In the unitary gauge we can take  $h^{\hat{a}}=(0,0,0,h)$, and
the matrix $U$ simplifies to
\begin{equation}
U=\left(\begin{array}{ccccc}
1 & 0 & 0 & 0 & 0\\
0 & 1 & 0 & 0 & 0\\
0 & 0 & 1 & 0 & 0\\
0 & 0 & 0 & \sqrt{1-s_h^2}& s_h\\
0 & 0 & 0 & -s_h & \sqrt{1-s_h^2} \end{array}\right)\,.
\label{mH11}
\end{equation}

The effect of the SUSY breaking is not felt at tree level by the colored fields $N_{L,R}$, $S_{L,R}$ mixing with the top.
We add to the SUSY scalar potential the SUSY breaking soft terms $V_{soft} =  V_{soft}^E+V_{soft}^C$ with 
\be
V_{soft}^E =  \widetilde m_{tL}^2 |\widetilde t_L|^2+  \widetilde m_{tR}^2 |\widetilde t_R|^2  \,, \ \ \ \ V_{soft}^C = \sum_{i=N_{L,R},S_{L,R}} \widetilde m_i^2 |\phi_i|^2  \,,
\label{EletR4}
\ee
and soft masses for the elementary gauginos of the SM gauge group, $\widetilde{m}_{1,2,3}$. We neglect the smaller soft mass terms radiatively induced by $W_0$ and
for simplicity we have not included $B$-terms.
Let us analyze the tree-level mass spectrum of the model. We fix the mass parameter $m_S = 0$, since all the states remain massive in this limit,\footnote{We checked that, if taken non-zero, its contribution to the potential does not change qualitatively the conclusions of our analysis.} and take $\lambda_L=\lambda_R=\lambda$, so that the composite superpotential enjoys a further ${\bf Z}_2$ symmetry (exchange of $L$ and $R$ fields), broken only by the mixing with SM fermions. 
We also assume all parameters to be real and positive. Before EWSB, the fermion mass spectrum in the matter sector is as follows.
A linear combination of fermions, to be identified with the top, is clearly massless. The $\SU(2)_L$ doublet with hypercharge $7/6$ contained in $N_{L,R}$ does not mix with other fields and has a mass equal to 
$M_{Q_{7/6}}=m_N$. The doublet with hypercharge $1/6$  mixes with $q_L$ and gets a mass  $M_{Q_{1/6}} = \sqrt{m_N^2+\epsilon_L^2}$.  Two  $\SU(2)_L$ singlets get a  
a mass square equal to $M_{S_{\pm}}^2=1/2(m_N^2+\epsilon_R^2+\lambda^2 f^2\pm \sqrt{(m_N^2+\epsilon_R^2)^2+2 m_N^2 f^2\lambda^2})$.
The scalar spectrum is analogous, with the addition of a shift given by the soft masses (\ref{EletR4}).
After EWSB, the top mass is 
\be
M_{top} = \frac{\epsilon_L \epsilon_Rf \lambda }{\sqrt{2}\sqrt{m_N^2+\epsilon_L^2}\sqrt{2\epsilon_R^2+f^2\lambda^2} }s_h \sqrt{1-s_h^2} =\frac{\epsilon_L \epsilon_Rf^2 \lambda^2 }{2\sqrt{2}M_{Q_{\frac 16}}M_{S_+}M_{S_-}}s_h 
\sqrt{1-s_h^2} 
\,.
\label{eq:mtoptRelemSinglet}
\ee

The gauge sector contains the SM vector superfields $w^{(0)}$ and $b^{(0)}$ and the chiral superfields $q_a$, $\psi_a$ and $Z$.
For simplicity, we neglect all soft mass terms in this sector, but the SM gaugino masses.
Regarding the fermion spectrum, the $\SO(4)$ fourplets $q_n$ and $\psi_n$ ($n=1,2,3,4$) get a Dirac mass $m_\psi$. A linear combination of $\psi_5$ and $Z$, we call it $p_5$, gets a Dirac mass, together with $q_5$, $\sqrt{2(f^2h^2+m_\psi^2)}$. The orthogonal combination of $\psi_5$ and $Z$ ($\chi_5$) is massless being the goldstino associated to the spontaneous breaking of SUSY.
In the scalar sector, ${\rm Re}\,q_n$ are identified as the pNGB Higgs, while ${\rm Im}\,q_n$ get a mass $\sqrt{2}m_\psi$. These two are the mass eigenstates of the two Higgs doublets $H_u$, $H_d$ introduced in section \ref{sec:HiggsMass}.
The partners of $\psi_n$ and $p_5$ will get the same mass as the fermions while the partner of the goldstino $\chi_5$ is a pseudo-modulus, whose VEV is undetermined at the tree-level.
This field is stabilized at the origin by a one-loop induced potential. Its detailed mass depends on the ratio $\mu^2 h^2/m_\psi^2$. In the region defined in the next subsection, its mass is of order $m_\psi h/(2\pi)\sim 50\div 70$ GeV.
The real and imaginary parts of $q_5$ have masses $\sqrt{2}fh$ and $\sqrt{2(f^2h^2+m_\psi^2)}$, respectively.

Let us discuss the possible values of the parameters of the model. 
Demanding $M_{top}$ to be around 150 GeV at the TeV scale gives a lower bound on the smallest possible value of the Yukawa coupling $\lambda$ at the scale $f$, obtained by taking
$\epsilon_{L,R}\rightarrow \infty$ in eq.(\ref{eq:mtoptRelemSinglet}):
\be
\lambda_{min}(f) \gtrsim 1.2\,.
\label{lambdamin}
\ee
An upper bound on $\lambda$ is found by looking at its RG running. 
For $h\ll 1$, the Yukawa coupling $\lambda$ is UV free for $\lambda(f)\lesssim 0.9$ and develops a Landau pole for higher values.
Demanding that the pole is at a scale greater than $4\pi f$ gives the upper bound:
\be
\lambda_{max}(f) \lesssim 1.7\,.
\label{lambdaMax}
\ee
Putting all together, we see that the maximum scale for which the model is trustable and weakly coupled is obtained by taking $\lambda=\lambda_{min}$, in which case we get
a Landau pole at around $300f$. This limiting value is never reached in realistic situations, but Landau poles as high as 100$f$ can be obtained.
The current bound on the top partner with 5/3 electric charge puts a direct lower bound on $m_N$ \cite{Chatrchyan:2013wfa}:\footnote{This bound can be applied directly only if the lightest top partner is this one with $Q = 5/3$, in which case it decays in $t W^+$ with $BR \simeq 100\%$. For lower values of the BR the bound is weaker. We take a conservative approach and use the bound as a constraint on the mass of this particle.}
\be
	M_{Q_{7/6}} = m_N  \gtrsim 800 \; {\rm GeV}\,.
	\label{Bound76}
\ee
Demanding a value for $\lambda$ as close as possible to the minimum value (\ref{lambdamin}), the top mass (\ref{eq:mtoptRelemSinglet}) favours regions in parameter space where $t_L$ and $t_R$ strongly mix with the composite sector:
$\epsilon_{L,R}\gg m_N$.

\subsection{Higgs Mass and Fine-Tuning Estimate}

As we have already remarked, when $V_{soft}^C$ is $\SO(5)$ invariant, 
the one-loop matter contribution to the Higgs potential is RG invariant and finite. 
Since the explicit form of $\beta_{matter}$ is quite involved, there is not a simple analytic expression for the Higgs mass valid in all the parameter space.
In particular an expansion for small values of $\epsilon_{L,R}$ is never a good approximation because, as explained, these mixing should be taken large.

The region of parameter space which realizes EWSB with $\xi = 0.1$ and gives $M_H=126$ GeV is essentialy unique.
In most of the parameter space $\gamma_{gauge} $ and $\gamma_{matter}$ are both positive and bigger than $\beta_{matter}$, and no tuning is possible to obtain the right value of $\xi$. The only region where $\gamma_{gauge} < 0$ is found for $\widetilde{m}_g, m_\psi \lesssim f$ where, however, the size of $\gamma_{gauge}$ is smaller than the natural size of $\gamma_{matter}$, eq.\eqref{mH2}. The bound \eqref{eq:boundmHSQ} forces $\widetilde{m}_H^2$ to be negligibly small.
From these considerations we see that $\gamma_{matter}$ has to be tuned in order to become smaller than its natural value. The requirement of perturbativity up to $\Lambda = 100 f$ fixes $\lambda(f) \simeq 1.3$. Regarding $m_N$, a lower bound is given by eq.\eqref{Bound76} while an upper bound is given from the fact that, increasing $m_N$ requires a higher value of $\epsilon_L$ in order to reproduce $M_{top}$, see eq.\eqref{eq:mtoptRelemSinglet}, and, as a consequence, $\gamma_{matter}$ increases, which is the opposite of what it is necessary to get $\xi$. This forces $m_N \sim f$, near its lower bound. Reproducing $M_{top}$ fixes also $\epsilon_L, \epsilon_R \gg f$. The Higgs mass is not sensitive to the stop soft masses $\widetilde{m}_{t_{L,R}}$ and its correct value is found for composite soft masses $\widetilde{m} \sim 3.5 f$, taken all equal. Finally, in order to fix $\xi = 0.1$,  $ \widetilde{m}_{t_L}$ and $\widetilde{m}_{t_R}$ have to be tuned in the region $\widetilde{m}_{t_L} \gg \widetilde{m}_{t_R} \sim \widetilde{m}$.

An approximate analytic formula for $M_H^2$ in this region is obtained by expanding for $\lambda f\ll  m_N, \widetilde m\ll \epsilon_{L,R}$, where $\widetilde m$ is a common universal soft mass term (the last one is a good approximation because $M_H$ does not depend on the stop soft masses). 
In this limit we get
\be
M_H^2 \simeq \frac{N_c}{2\pi^2}M_{top}^2\frac{M_{top}^2}{v^2}\bigg(5 \log\Big(\frac{\widetilde m^2}{\lambda_{top}^2 f^2}\Big)+ 4x\log \Big(\frac{x}{1+x}\Big)
+\frac 12-4\log 2\bigg)\,,
\label{Mh2}
\ee
where
\be
x=\frac{\widetilde m^2}{m_N^2}\,.
\ee
It is immediate to see that for values of $\widetilde m \gtrsim  m_N \sim f$,\footnote{We have numerically checked that the range of applicability of eq.(\ref{Mh2}) extends to the region with $m_N\sim f$.} a 126 GeV Higgs is reproduced.

Let us briefly discuss the fine-tuning. We define it here as the ratio between the value of $\xi$ we want to achieve and its natural value given by (\ref{xi0}) in absence of cancellations.
This is a crude definition, but it has the advantage to estimate the actual fine-tuning provided by cancellations rather than the sensitivity,  without the need to worry about  possible generic sensitivities.
The electroweak scale is determined by eq.(\ref{xi0}). As argued above most of the tuning arises within the matter sector.
We can then neglect $\gamma_{gauge}$ and determine the expected value of $\xi$ by comparing $\gamma_{matter}$ and $\beta_{matter}$. 
We get\footnote{As explained below eq.(\ref{Vsusy0}), the limits $s_h\rightarrow 0$ and $\widetilde  m\rightarrow 0$ do not commute. As a result, $\beta_{matter}$ in eq.(\ref{eq:gammabetatRelem}) does not vanish for  $\widetilde  m\rightarrow 0$.}
\be
\gamma_{matter}\sim  \frac{N_c}{8\pi^2}\lambda_{top}^2 f^2\widetilde m^2\,, \ \ \ \ \ 
\beta_{matter} \sim  \frac{N_c}{8\pi^2}\lambda_{top}^4 f^4\,.
\label{eq:gammabetatRelem}
\ee
\begin{figure}[t]
\begin{center}
		\includegraphics[width=110mm]{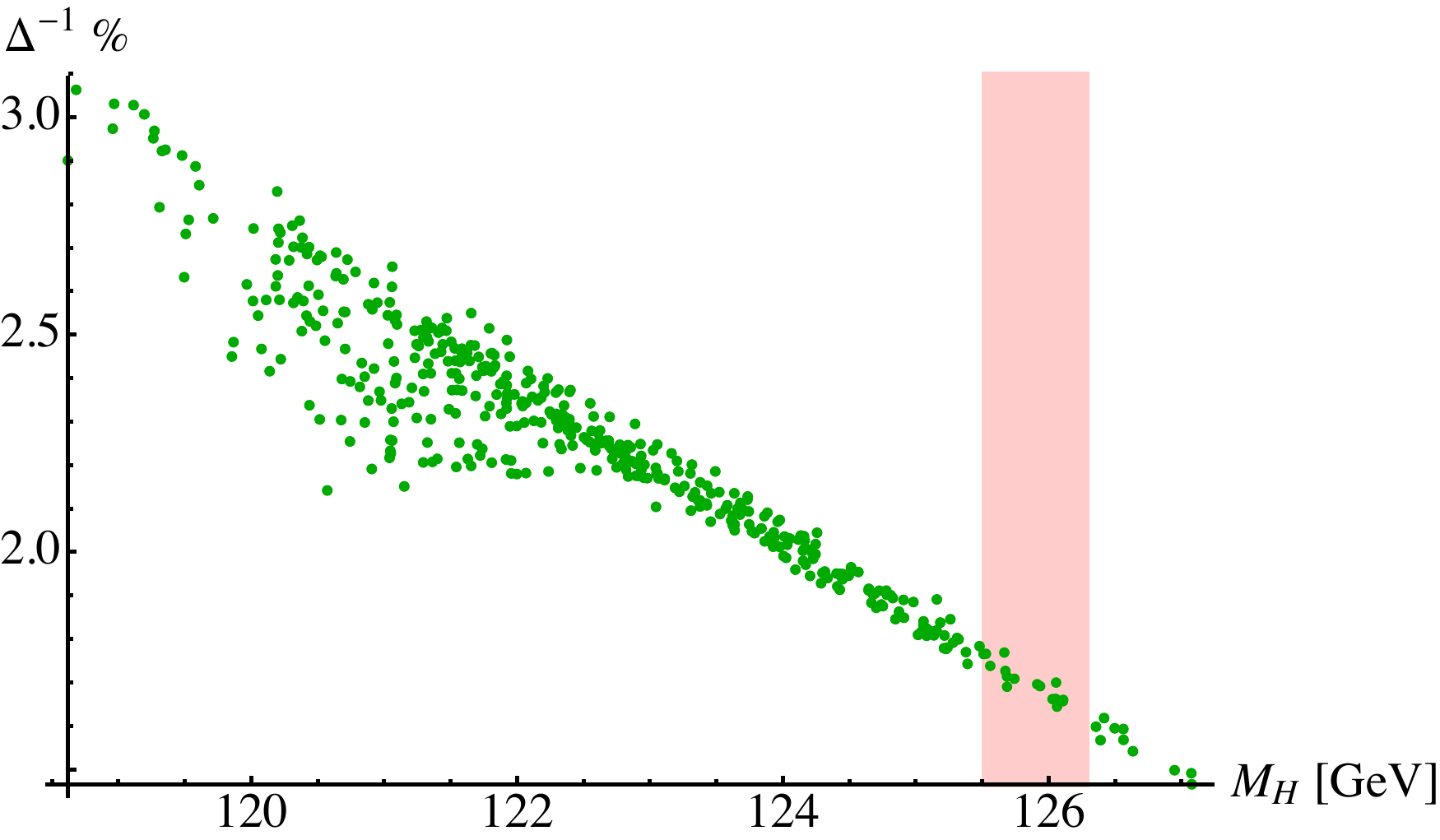}
\end{center}
\caption{ \small Fine-tuning of the minimal model, (in \%) as a function of the Higgs mass for $\xi \simeq 0.1$. In this scan we fixed $\lambda(f) = 1.29$ and $h(f) = 0.44$, so that both $\lambda$ and $h$ reach a Landau pole  at the same scale $\Lambda \sim 100 f$, $m_N = 1.2 f$ and picked randomly $\epsilon_L \in [8.5f, 10f]$, $m_{\psi} \in [0.7 f, 1.5 f]$, $\widetilde{m}_g \in [0.5f, f]$, $\widetilde{m} \in [2.5f, 4.5f]$, $\widetilde{m}_{t_L} \in [4.5f, 6.5f]$, $\widetilde{m}_{t_R} \in [f, 3f]$ and $\widetilde{m}_H^2$ within the bound \eqref{eq:boundmHSQ}. We fixed $M_{top}$ by solving for $\epsilon_R$ and then selected points with $\xi \simeq 0.1$. The pink strip represents the current Higgs mass 1$\sigma$-interval as reported in ref.~\cite{PDG2013}.}
\label{fig:FT1}
\end{figure}
The fine-tuning can then be written as\footnote{Another possible source of fine-tuning might arise from the origin of the scale $f$ as the cancellation of the two terms in eq.(\ref{fDEF}). In the region of interest no significant cancellation occurs and we neglect this effect.}
\be
	\Delta \sim \frac{\widetilde m^2}{f^2}\frac{1}{\xi} \,, 
\label{EletR14}
\ee
and is always higher than the minimum value $1/\xi$. From eq.(\ref{Mh2}) we see that $M_H$ grows with $\widetilde m$ in the region of interest and hence we expect a linear increase of $\Delta$ with the Higgs mass.

In order to check these considerations we performed a parameter scan in the restricted region described at the beginning of the section. We fixed the top mass by solving for $\epsilon_R$ and then obtained the minimum of the potential and the Higgs mass from the full one-loop expression of eq.\eqref{eq:1loop_pot}.
We report in fig.\ref{fig:FT1} a plot of the fine-tuning computed using the standard definition of ref.~\cite{Barbieri:1987fn} as a function of the Higgs mass.
As can be seen, we obtain $\Delta^{-1} \sim 2 \%$ for $M_H = 126$ GeV, in reasonable agreement with the rough estimate (\ref{EletR14}).

Let us now discuss the spectrum of new particles. In this region, the electroweak gauginos are relatively light, $\widetilde{m}_g \lesssim f \sim 800$ GeV and the two higgsino  doublets (from $\psi_n$ and $q_n$) have also a mass $m_\psi \sim 800$ GeV. The stops and their partners are heavy, above 2 TeV, while the fermion top partners are usually below the TeV, the lightest being the singlet with $Q = 2/3$ and a mass $M_{S_-} \simeq 660$ GeV.\footnote{The recent CMS analysis \cite{Chatrchyan:2013uxa} rules out charge 2/3 top partners below $\sim $ 700 GeV. A careful phenomenological analysis should be performed to check if the model with the benchmark parameters taken is ruled out or not. Slightly decreasing $\xi$ or the scale of the Landau pole are two possible solutions
to increase the mass of $M_{S_-}$ beyond 700 GeV.} The gluinos do not contribute to the Higgs potential at one loop, therefore they can be taken heavy (above the experimental bounds) without increasing the fine-tuning.

\section{Model with Vector Resonances}
\label{sec:tREVect}

The model we consider in this section is essentially the IR  effective description of the UV model with elementary $t_R$  introduced in ref.~\cite{Caracciolo:2012je}. 
The symmetry breaking pattern is
\be
 \SO(5) \times \SO(4)_2\times \U(1)_X  \rightarrow  \SO(4)_D \times \U(1)_X \, ,
\label{eq:Gbreak}
\ee
where $\SO(4)_2$ is gauged and $\SO(4)_D$ is the diagonal subgroup of $\SO(4)_2$ and $\SO(4)_1 \subset \SO(5)$. The electroweak gauge group is $\SU(2)_L^0 \times \U(1)_Y^0 = G_{SM} \subset \SO(4)_1\times \U(1)_X$. This group structure introduces a partial compositeness mechanism also for the electroweak vector multiplets, in close analogy to what happens in non-SUSY CHM in presence of vector resonances.

The superpotential is 
\be
W = \sum_{i=L,R} (\epsilon_i \xi^a_i N_i^a + \lambda_i X_i^n q^a_n N_i^a )+ m_N N_L^a N_R^a + m_X  X_L^n X_R^n+ W_0(Z,q)\,,
\label{WtEVect}
\ee
where $N_{L,R}$ and $X_{L,R}$ are colored fields in the  $({\bf 5},{\bf 1})$, $({\bf 1}, {\bf 4})$ of $\SO(5) \times \SO(4)_2 $, respectively, and $q$ is a color-singlet in the $({\bf 5},{\bf 4})$ ($a=1,\ldots,5$, $n=1\ldots,4$).
The spurions $\xi_{L,R}$ are taken as in eq.(\ref{EletR2}).
The superpotential term $W_0$ reads
\be
	W_0 = h\Big( q^n_a Z_{ab} q^n_b -  \frac{f^2}{2}  Z_{aa} \Big),
	\label{W0}
\ee
where $Z$ is a field in the symmetric ${\bf 14}\oplus {\bf 1}$ of $\SO(5)$. 
The field $q$ acquires a VEV  
\be
\langle q_{a}^n \rangle = \frac{1}{ \sqrt{2}} f  \delta_a^n 
\label{qVEV}
\ee
in its scalar component. In this vacuum  the symmetry group $G$ is broken as in eq.(\ref{eq:Gbreak}) and SUSY is broken by the rank condition \cite{Intriligator:2006dd}. 
The spontaneous breaking of SUSY is necessary to give mass to the fermions  and the scalars inside $q$, which are the higgsinos and the scalar partners of the NGB Higgs. 
The effect of the SUSY breaking is not felt at tree level by the colored fields mixing with the top, i.e. $N_{L,R}$ and $X_{L,R}$.
Like in the previous model, eq.(\ref{EletR4}), we add SUSY breaking soft terms for $t_{L,R}$, $N_{L,R}$ and $X_{L,R}$, neglecting $B$-terms and the smaller soft mass terms radiatively induced by $W_0$.

Ten NGB's result form the breaking (\ref{eq:Gbreak}):  four $h^{\hat a}$ and six $\pi^A$ transforming in the  ${\bf 4}$  and  $({\bf 3},{\bf 1})\oplus({\bf 1},{\bf 3})$ of  $\SO(4)_D$, respectively.
The $h^{\hat a}$ correspond to the four real Higgs components while the extra unwanted $\pi^A$'s are eaten by the $\SO(4)_2$ gauge fields. In the unitary gauge $h^{\hat{a}}=(0,0,0,h)$, $\pi^A = 0$,
we have
\be
	q_a^n = U_{ab} \tilde q_b^n\,,
	\label{mH10a}
\ee
where $\tilde q$ encode the non-NGB degrees of freedom of $q$ and $U$ is the matrix (\ref{mH11}).

The matter sector includes the fields $N_{L,R}$, $X_{L,R}$, and the spurions $\xi_{L,R}$, while the 
 gauge sector include the SM gauge fields, the fields $Z$ and the non-NGB components $\tilde q$ of $q$.
All the parameters in the superpotential (\ref{WtEVect})  are taken positive and we neglect $m_X$.
We also assume $\lambda_L=\lambda_R=\lambda$ for simplicity.\footnote{In the model of ref.~\cite{Caracciolo:2012je}, these choices are dynamically realized. For example, $m_X$ would correspond to a mass term
for the dual magnetic quarks and is not generated in the superpotential.}
Before EWSB, the fermion mass spectrum in the matter sector is as follows.
One linear combination of fermions, to be identified with the top, is clearly massless. The $\SU(2)_L$ doublet with hypercharge $7/6$ and $1/6$ contained in $N_{L,R}$ and $X_{L,R}$
have a mass square $M_{Q_i\pm}^2=1/2(\alpha_i+\lambda^2 f^2 \pm \sqrt{\alpha_i^2+2 m_N^2 f^2\lambda^2})$,
where $\alpha_{7/6}=m_N^2$ and $\alpha_{1/6}=m_N^2+\epsilon_L^2$. The  $\SU(2)_L$ singlet has a mass $M_{S}=\sqrt{m_N^2+\epsilon_R^2 f^2}$.
The scalar spectrum is analogous, with the addition of a shift given by the soft masses (\ref{EletR4}).
After EWSB, the top mass is 
\be
M_{top} = \frac{\epsilon_L \epsilon_Rf \lambda }{\sqrt{2}\sqrt{m_N^2+\epsilon_R^2}\sqrt{2\epsilon_L^2+f^2\lambda^2} }s_h \sqrt{1-s_h^2} =
\frac{\epsilon_L \epsilon_R \lambda^2 f^2}{2 \sqrt{2} M_{Q_{\frac16+}} M_{Q_{\frac16-}} M_S} s_h \sqrt{1-s_h^2} \,.
\label{EletR5}
\ee

The gauge sector contains the chiral superfields $q$, $Z$, the vector superfields $\rho$ in the adjoint of  $\SO(4)_2$, in addition to the usual SM vector superfields $w^{(0)}$ and $b^{(0)}$.
For simplicity, we neglect all soft mass terms in this sector, but the SM gaugino mass terms, namely
the $\SO(4)_2$ gaugino mass terms and scalar mass terms for the $Z$ and $q$ fields, as well as $B$-terms.
In this limit, all the fields of the gauge sector, but $q_5^n$ and $Z_{5n}$, do not feel at tree-level the SUSY breaking induced by the $F$-term of $Z_{55}$
and have a SUSY spectrum.
The chiral multiplets $(q^m_n+q^n_m)/\sqrt{2}$ and $Z_{mn}$ combine and get a mass $\sqrt{2}h f$;  
the chiral fields $(q_n^m-q_m^n)/\sqrt{2}$ combine with the $\SO(4)_2$ vector multiplets into  a massive vector super-field with (up to ${\cal O}( g_{SM} / g_\rho$) effects)
\be
M_\rho = g_\rho f\,,
\ee 
where $g_\rho$ is the coupling of the $\SO(4)_2$ gauge theory. The scalar field $Z_{55}$  is stabilized at the origin at the radiative level and gets a mass $\simeq h f/\pi$.
Its fermion partner (a complete singlet) is massless at this order, being the goldstino. 
The higgsinos $\psi_{q_5}$ and $\psi_{Z_{5m}}$ mix and get a Dirac mass $h f$, the scalars $Z_{5m}$ also get a mass $h f$. 
The scalars in the $\bf 4$ of $\SO(4)_2$, $q_5^m$, behave as in the model of section \ref{sec:tRE}:  ${\rm Re}\, q_5^m$ are massless at tree-level being the pNGB Higgs doublet while ${\rm Im}\, q_5^m$ get a mass $\sqrt{2} hf$.

Let us discuss on the possible range of the parameters of the model. 
The top mass gives the same lower bound found in eq.(\ref{lambdamin}) for $\lambda(f)$.
Demanding vector resonances masses above 2 TeV fixes, for $f\simeq 800$ GeV, $g_\rho(f)\simeq 5/2$ and a Landau pole for $g_\rho$ at $\Lambda \simeq 4\pi f$.
We can determine the values of $\lambda$ and $h$  at the scale $f$ by the requirement that they reach a Landau pole at the same scale $\Lambda$ where $g_\rho$ blows up:
\be
\lambda(f) \simeq 2\,, \ \ \ \ \ h(f)\simeq 0.9 \ \,.
\label{PoleVect}
\ee
Taking $\lambda\simeq 2$ gives an upper bound on $m_N$, which comes from the bound (\ref{Bound76}).  We get
\be
m_N  \lesssim 1.2 f.
\label{Bound76V2}
\ee

\subsection{Higgs Mass and Fine-Tuning Estimate}

We performed a numerical study of the Higgs potential by fixing $\lambda$ and $h$ as in eq.(\ref{PoleVect}), one mass mixing
parameter by demanding $M_{top}({\rm TeV})\simeq 150$ GeV and scanning over the remaining parameters. 
Even tough the analysis is numerical, it is possible to get an understanding of the preferred region in parameter space through some considerations. 
Unlike the model with no vector resonances discussed in section \ref{sec:tRE}, now $\gamma_{gauge}$ is negative and increases with the gaugino masses (see eq.\eqref{gammaMatGauge}), and it can be tuned against $\gamma_{matter}$ in order to obtain $\xi=0.1$ without the need to tune the latter to unnaturally small values. The correct Higgs mass can be obtained by raising either $\epsilon_L$ and/or the composite soft masses, taken all equal to $\widetilde{m}$. A smaller tuning is achieved by taking $\epsilon_L \gg \epsilon_R \sim f$ and $\widetilde{m} \sim f$.
In order to further decrease the tuning (i.e. to decrease $\gamma_{matter}$ and, therefore, the needed gaugino masses) the preferred stop soft masses are $\widetilde{m}_{t_R} > \widetilde{m}_{t_L} \sim \widetilde{m}$. Finally, in order to get $\xi = 0.1$ one can tune $\gamma_{gauge}$ choosing the right wino and bino soft masses. Their typical size is $\widetilde{m}_g \sim 3f$.
Unfortunately we have not found a  simple analytic expression for the Higgs mass in this region of parameter space.

\begin{figure}[t]
\begin{center}
		\includegraphics[width=110mm]{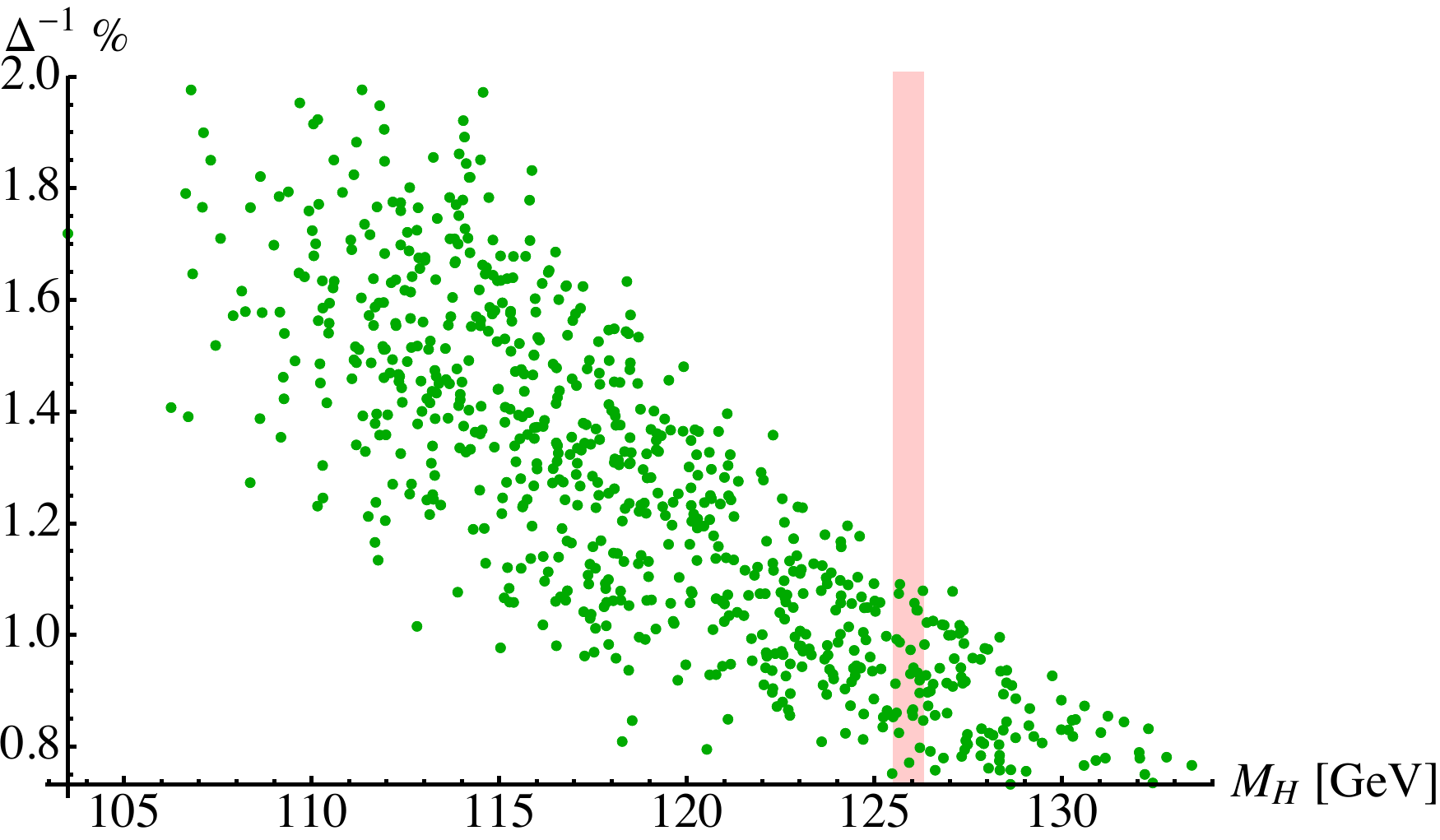}
\end{center}
\caption{ \small Fine-tuning of the model with vector resonances (in \%) as a function of the Higgs mass for $\xi \simeq 0.1$. We fixed $\lambda(f) = 2.1$, $h(f)=0.9$ and $g_\rho(f) = 2.6$ so that $\lambda$, $h$ and $g_\rho$ reach a Landau pole at the same scale $\Lambda \sim 4 \pi f$, $m_N = 1.2$ and picked randomly $\epsilon_L \in [3f, 7f]$, $\widetilde{m}_g \in [2.5f, 4.5f]$, $\widetilde{m} \in [f, 2.5f]$, $\widetilde{m}_{t_L} \in [f, 4f]$, $\widetilde{m}_{t_R} \in [2f, 5f]$ and $\widetilde{m}_H^2$  within the bound \eqref{eq:boundmHSQ}. We fixed $M_{top}$ by solving for $\epsilon_R$ and then selected points with $\xi \simeq 0.1$.  The pink strip represents the current Higgs mass 1$\sigma$-interval as reported in ref.~\cite{PDG2013}.}
\label{fig:FT2}
\end{figure}

Let us briefly discuss the necessary amount of fine-tuning.  In the above region of interest,  most of the tuning arises from a cancellation occurring between $\gamma_{gauge}$ and $\gamma_{matter}$.
In fact, the large gaugino mass is needed only to get the correct EWSB pattern.
For $m_N\sim \epsilon_R\sim f$, $\widetilde m_g\gg f, M_\rho$, $\widetilde m>f$   and $g\ll g_{\rho}$, we have 
\be
\begin{split}
\gamma_{gauge} & \sim -\frac{3g^2  f^2}{32\pi^2}\widetilde m_g^2\Big(\log\frac{\widetilde m_g^2}{f^2}-1\Big)\, ,  \\
\gamma_{matter} & \sim \frac{N_c f^2}{16\pi^2} \widetilde m^2 \Big(\log\frac{\epsilon_L^2}{\widetilde m^2}+1\Big)\, .
\end{split}
\label{gammaMatGauge}
\ee
We see that $\gamma_{gauge}$ and $\gamma_{matter}$ have opposite signs and fine-tuning is possible.
In the region of interest, one can grossly estimate $\beta_{matter}  \sim N_c f^4/(16\pi^2)$. 
Using eq.(\ref{xi0}), we roughly get 
\be
\Delta \sim \frac{\widetilde m^2}{f^2}\frac{1}{\xi} \,, 
\label{FT}
\ee
and coincides with the estimate (\ref{EletR14}) done for the model in section \ref{sec:tRE}. Since $M_H$ increases with $\widetilde m$, we expect again the fine-tuning to grow with $M_H$.

We report in fig.\ref{fig:FT2} a plot of the fine-tuning, computed using the standard definition of 
ref.~\cite{Barbieri:1987fn}, as a function of the Higgs mass for our parameter scan.
As can be seen, the tuning is of the order $\Delta^{-1} \sim 1 \%$ and is in rough quantitative agreement with eq.(\ref{FT}).

Turning to the new-physics spectrum, in this model we find heavy electroweak gauginos, $\widetilde{m}_g\simeq 3$ TeV and heavy higgsinos, $M_{higgsinos} \simeq 700-800$ GeV. The stops and their partners are usually above the TeV, while the fermion top partners are below, the lightest having $Q=5/3$ and a mass $M_{Q_{5/3}} \simeq 800-900$ GeV. As for the previous model, the gluinos can be decoupled without increasing the fine-tuning.

\section{Conclusions}
\label{conclusions}

In this paper we have considered a possible framework for models based on SUSY and a pNGB Higgs with partial compositeness,
focusing on the Higgs potential and the expectation for the Higgs mass. 
We have constructed two specific models of this form where the top quark is elementary and the Higgs arises as a pNGB of a linearly realized $\SO(5)\rightarrow \SO(4)$ breaking pattern.
In both models the matter fields in the composite sector are taken in the fundamental or singlet representations of $\SO(5)$. En passant, we have generally shown that, independently of SUSY,
a scenario where $t_R$ is fully composite is disfavoured because it tends to give a too light Higgs.

The first model, with no vector resonances, can be seen as an elementary linear completion of a pNGB Higgs, while the second one
is more properly interpreted as an IR description of a strongly coupled gauge theory in terms of composite resonances, although none of the above results depends on the details of the microscopic completion: in particular, the Higgs potential is calculable and the only logarithmic UV sensitivity is introduced by the gaugino soft mass terms.

In both models the parameter space is quite constrained, especially from the appearance of Landau poles for certain Yukawa couplings and from the top mass constraint.
Requiring a reasonable range of perturbativity and reproducing the top mass implies fermion top partners in both models with a mass around the compositeness scale $f$ ($\sim 800$ GeV for reasonably natural theories), 
independently of the Higgs mass.  These are the lightest exotic colored states, while the pNGB nature of the Higgs allows to decouple stops and gluinos with no fine-tuning issues.
The minimal model with no vector resonances predicts electroweak gauginos and higgsinos with a mass near $800$ GeV, while in the model with vectors electroweak gauginos are heavier, usually above the TeV scale.

From the above discussion, we expect that the most sensitive channels for the discovery, or exclusion, of these models will be fermion top partner searches. In particular, the bound on the $Q=5/3$ top partner mass already puts severe constraints on the parameter space of our models.
A prominent phenomenological feature, relevant for superpartners phenomenology, is the possible presence of an R-parity which, in particular, makes the lightest supersymmetric particle  stable. The deviations in the Higgs couplings to gauge bosons and top quark are of order $\xi$ and are analogous to those of non-SUSY CHM.

\section*{Acknowledgments}

We  thank Brando Bellazzini, Roberto Contino, Nathaniel Craig, Riccardo Rattazzi, Andrea Romanino, Giovanni Villadoro and Andrea Wulzer  for useful discussions. M.S. thanks the Galileo Galilei Institute for Theoretical
Physics for hospitality during the completion of this work.
D.M. thanks the Theory Division at CERN for hospitality and the contract PITN-GA-2009-237920 \emph{UNILHC} for support during the completion of this work.

\appendix

\section{A More Accurate Parametrization of the Higgs Potential}
\label{app:delta}

For $s_h\ll 1$, the tree-level + one-loop potential  $V=V^{(0)}+V^{(1)}$ admits an expansion of the form
\be
V = - \gamma s_h^2 +\beta s_h^4 + \delta s_h^4 \log s_h+{\cal O}(s_h^6)\,.
\label{PotExpdeltaA}
\ee
The last non-analytic term cannot obviously be obtained by a Taylor expansion around $s_h=0$. It arises at the one-loop level and is due to the contribution of particles  whose mass vanishes for $s_h=0$.
In a naive expansion around $s_h=0$, its presence would be detected by the appearance of a spurious IR divergence in the coefficient $\beta$.
At first order in $\delta$, the non trivial minimum of the potential is found at
\be
\langle s_h^2 \rangle \equiv \xi = \xi_0 \Big(1-\frac{\delta}{4\beta}(1+2\log \xi_0) \Big)\,,
\ee
where
\be
\xi_0 = \frac{\gamma}{2\beta}
\label{xi0A}
\ee
is the leading order minimum for $\delta = 0$. The Higgs mass is given by
\be
M_H^2 =  \frac{8 \beta}{f^2} \xi_0 (1-\xi_0) + \frac{4\delta \xi_0}{f^2}\Big(1-\frac{\xi_0}2+\xi_0\log \xi_0\Big).
\ee
For $\xi_0\ll 1$ we get
\be
M_H^2 \simeq (M_H^0)^2 \Big(1+\frac{\delta}{2\beta}\Big),
\label{mHcorr}
\ee
where
\be
(M_H^0)^2 \simeq \frac{8\beta}{f^2} \xi_0
\label{mH0A}
\ee
is the leading order mass for $\delta = 0$. The Higgs mass squared formula reported in eq.(\ref{Mh2})  refers to eq.(\ref{mHcorr}), where the non-analytic term is included at linear order in $\delta$.

In the models we considered, the particles massless at $s_h=0$ are always the top in the matter sector and the $W$ and the $Z$ gauge boson in the gauge sector.
Correspondingly, the explicit form of $\delta=\delta_{gauge}+\delta_{matter}$ is universal and given by 
\be\begin{split}
\delta_{matter}  & = - \frac{N_c}{8\pi^2} \lambda_{top}^4 f^4\,, \\
\delta_{gauge}  = & \frac{3f^4 (3g^4+2 g^2 g^{\prime 2}+g^{\prime 4})}{512\pi^2} \,,
\label{deltaUni}
\end{split}
\ee 
with $N_c=3$ the QCD color factor and  $M_{top}\equiv \lambda_{top} v$.

\section{Unitarization of $WW$ Scattering}

\label{WWscatte}

In theories where the Higgs is a pNGB the scattering amplitudes between the longitudinal polarizations of the $W$ and $Z$ bosons, and the Higgs itself, for energies higher than the Higgs mass grow quadratically with the energy, ${\cal A} \sim E^2 / f^2$, violating perturbative unitarity at a scale $\Lambda \sim 4 \pi f$. At this scale, or before, new degrees of freedom (in the form of either strong dynamics effects or new perturbative fields) must become important in the scattering to restore unitarity.
In the following we will see how the field content present in each model is exactly what is needed to restore perturbative unitarity of $WW$ scattering, as expected from linear models.

In the $\SO(5) \rightarrow \SO(4)$ coset, all $h^{\hat{a}} h^{\hat{b}}$ scattering amplitudes can be parametrized in terms of only two functions of the Mandelstam variables, $A(s,t,u)$ and $B(s,t,u)$ \cite{Contino:2011np}:
\be
	{\cal A}(h^{\hat{a}} h^{\hat{b}} \rightarrow h^{\hat{c}}h^{\hat{d}}) = A(s,t,u) \delta^{\hat{a}\hat{b}} \delta^{\hat{c}\hat{d}} + A(t,s,u) \delta^{\hat{a}\hat{c}} \delta^{\hat{b}\hat{d}} + A(u,t,s) \delta^{\hat{a}\hat{d}} \delta^{\hat{b}\hat{c}} + B(s,t,u) \epsilon^{\hat{a}\hat{b}\hat{c}\hat{d}}.
\ee
In our models, however, in the limit of zero SM gauging the gauge sector has a $P_{LR}$ symmetry which fixes $B(s,t,u) = 0$. The NGB contribution to the scattering is universal and given by
\be
	A_{NGB} (s,t,u)= \frac{s}{f^2}.
\ee
The possible contributions to NGB scattering can be obtained simply by group theory and the fact that bosonic states must be symmetric under the exchange of identical particles:
\be \begin{split}
	\text{$h^{\hat{a}}h^{\hat{b}}$ scattering: } & {\bf 4} \otimes {\bf 4} = ({\bf 1}; J=0) \oplus ({\bf 6}; J=1) \oplus ({\bf 9}; J=0) \ ,
	\label{eq:AllowedStatesScatt}
\end{split} \ee
where $J$ is the spin.

\subsection{Minimal Model $\SO(5) \rightarrow \SO(4)$}

In this theory the only NGB present are the four components of the Higgs doublet, $h^{\hat{a}}$, and there is no  gauge boson other than the SM ones.
The gauge sector of the model is a supersymmetrization of the liner $\sigma$-model presented in ref.~\cite{Barbieri:2007bh} and in the Appendix G of ref.~\cite{Contino:2011np}. The only term in the Lagrangian relevant to $WW$ scattering is the kinetic term of the real part of $q = (\phi + i \tilde{\phi}) / \sqrt{2}$, which takes a VEV $\langle \phi \rangle = (0,0,0,0,f)$.
Expliciting the NGB dependence as $\phi(x) = U(h^{\hat{a}}(x) ) \langle \phi \rangle \left(1+\frac{\eta(x)}{f}\right)$, where $\eta(x)$ is a real singlet scalar field with mass $M_\eta = \sqrt{2} h f$, we can write the Lagrangian as
\be
	\Lag_{kin} = \frac{1}{2} (\partial_\mu \eta)^2 - \frac{1}{2} M_\eta^2 \eta^2 + \frac{f^2}{4} \Tr \left[ d_\mu d^\mu \right] \left(1 + \frac{\eta}{f} \right)^2 \ ,
\ee
where we defined the Callan-Coleman-Wess-Zumino  structures \cite{Coleman:1969sm} $d_\mu^{\hat{a}} T^{\hat{a}} + E_\mu^a T^a = i U^\dag D_\mu U$ and $\nabla_\mu = \partial_\mu - i E_\mu$. The full NGB scattering amplitude in this theory can be written as
\be
	A(s,t,u) = \frac{s}{f^2} \left( 1 - \frac{s}{s - M_\eta^2} \right),
\ee
which evidently recovers perturbative unitarity for $\sqrt{s} \gg M_\eta$.

\subsection{Model with Vector Resonances}

This model has ten Goldstone bosons: six  $\pi^A$ in the adjoint representation of $\SO(4)_D$ and four $h^{\hat{a}}$ in the fundamental of $\SO(4)_D$. In the \emph{unitary gauge}, where the Goldstone bosons in the adjoint are eaten by the $\rho$ gauge bosons, the study of their scattering is shifted to the study of the $\rho \rho$ scattering.
With the aim to connect our study with previous bottom-up studies of the effect of resonances in $WW$ scattering in CHM and their phenomenology at the LHC \cite{Contino:2011np}, in the following we will concentrate only on the study of the scattering amplitudes among the four NGBs which form the Higgs doublet.

All contributions to NGB scattering, see eq.\eqref{eq:AllowedStatesScatt}, come from the kinetic term of the fields in the multiplet which takes a VEV triggering the spontaneous symmetry breaking, in our case the real components of $q_a^n$:
\be
	\Lag = \left| D_\mu q_a^n \right|^2 = \left| i U^\dagger D_\mu q_a^n \right|^2 =  \left|  i \nabla_\mu \tilde q + d_\mu \tilde q - g_\mu \tilde q \rho_\mu \right|^2 ,
	\label{eq:kineticLagrangianNGB}
\ee 
where we used eq.\eqref{mH10a} to render explicit the NGB dependence.
The fields $\tilde q_a^n$ transform under the unbroken group $\SO(4)_D \sim SU(2)_L \times SU(2)_R$ in the representations
\be
	\tilde q_a^n : \quad \bf{1 \oplus 9 \oplus 6 \oplus 4} = \bf{(1,1) \oplus (3,3) \oplus ( \: (1,3) \oplus (3,1) \:) \oplus (2,2)}\,.
\ee
Its decomposition in terms of component fields is
\be
	\tilde q_a^n (x) = \left(\frac{f}{\sqrt{2}} + \frac{\eta(x)}{2}\right) \delta_a^n + \Delta^{A_LB_R}(x) (2 T^{A_L} T^{B_R})_a^n + \frac{\tilde q_\rho^A (x)}{\sqrt{2}} (T^A)_a^n + i \frac{\tilde q_5^n (x)}{\sqrt{2}} \delta_{a5},
\ee
where the singlet $\eta$ and the symmetric traceless $\Delta$ are complex, while the antisymmetric $\tilde q_\rho$ and the fundamental $\tilde q_5$ are real fields.
From eq.\eqref{eq:AllowedStatesScatt} we see that the only states which can contribute to the scattering are the singlet ($\eta$), the gauge bosons ($\rho$) and the symmetric ($\Delta$). Since we are interested only in the tree-level contribution to the scattering amplitude, we can study them separately.

Let us start with the singlet $\eta = (\eta_1 + i \eta_2) / \sqrt{2}$. Setting $\Delta$ and $\rho_\mu$ to zero in eq.\eqref{eq:kineticLagrangianNGB} one can arrive easily to the Lagrangian\footnote{Here we also used that $\delta^n_c (T^A T^B)_{cd} \delta_d^n = \delta^{AB}$, $\delta^n_c (T^{\hat{a}} T^{\hat{b}})_{cd} \delta_d^n = \delta^{\hat{a} \hat{b}} / 2$ and  $\delta^n_c (T^A T^{\hat{b}})_{cd} \delta_d^n = 0$, where $T^A$ and $T^{\hat{a}}$ are, respectively, the unbroken and broken generators of $\SO(5) \rightarrow \SO(4)$.}
\be
	\Lag \supset \abs{\partial_\mu \eta}^2 + \frac{1}{2} \Tr \left[ d_\mu d^\mu \right] \abs{\mu + \frac{\eta}{2}}^2 = \frac{1}{2}\left(  (\partial_\mu \eta_1)^2 +  (\partial_\mu \eta_2)^2 \right) + \frac{f^2}{4} \Tr \left[ d_\mu d^\mu \right] \left(1 + \frac{\eta_1}{f} + \frac{\eta_1^2 + \eta_2^2}{4 f^2} \right).
\ee
In the parametrization of ref.~\cite{Contino:2011np} it is easy to recognize $a_{\eta_1} = \frac{1}{2}$, $a_{\eta_2} = 0$ and  $b_{\eta_1} = b_{\eta_2} = \frac{1}{4}$.
From this we obtain the contribution of the $\eta$ to the $hh$ scattering amplitude:
\be
	A_\eta(s,t,u) = - \frac{1}{4} \frac{s}{f^2} \frac{s}{s - M_\eta^2},
\ee
where $M_\eta = \sqrt{2} h f$.
Setting to zero the scalars $\Delta$ and $\eta$ we can obtain the contribution from the vector $\rho_\mu$. The Lagrangian can be written as
\be
	\Lag \supset \frac{f^2}{4} \Tr \left[ d_\mu d^\mu \right] + \frac{f^2}{2} \Tr \left[ (g_\rho \rho_\mu - E_\mu)^2 \right],
\ee
recognizing that, in the notation of ref.~\cite{Contino:2011np},  $a_\rho = 1$.
The contribution to the scattering amplitude which grows with the energy is
\be
	A_\rho(s,t,u) = - \frac{3}{2} \frac{s}{f^2}.
\ee
The scalar $\Delta = (\Delta_1 + i \Delta_2) / \sqrt{2} $ is a complex field in the $\bf{(3,3)}$ of $\SO(4)$.
Its Lagrangian can be written as
\be
	\Lag = \sum_{i=1,2} \left\{ \frac{1}{2} \Tr[(\nabla_\mu \Delta_i)^2] -  \frac{M_{\Delta_i}^2}{2} \Tr[\Delta_i^2] + a_{\Delta_i} f \Tr[\Delta d_\mu d^\mu] + \ldots \right\},
\ee
where, in components, $\Delta_i = \Delta^{A_LB_R}_i(x) (2 T^{A_L} T^{B_R})_a^b$ and where the dots represent terms not relevant for $WW$ scattering. 
In our case $a_{\Delta_1} = 1$, $a_{\Delta_2} = 0$ and $M_{\Delta_1} = M_{\Delta_2} = \sqrt{2} h f$.
The scattering amplitude is given by
\be
	A_\Delta(s,t,u) = \frac{(a_{\Delta_1}^2 + a_{\Delta_2}^2)}{4} \left( \frac{s}{f^2} \frac{s}{s - M_\Delta^2 } - 2 \frac{t}{f^2} \frac{t}{t - M_\Delta^2 } - 2 \frac{u}{f^2} \frac{u}{u - M_\Delta^2} \right).
\ee
For energies larger than the masses of these resonances we have
\be
A_{tot}(s,t,u) =A_{NGB} (s,t,u)+A_\eta(s,t,u)+A_\rho(s,t,u)+A_\Delta(s,t,u)  \, {\simeq}\, \text{const}.
\ee
We see that, as expected, the exchange of heavy resonances restores unitarity before the scattering amplitude becomes non perturbative.


\end{document}